\documentclass[10pt,a4paper,twoside]{article}
\usepackage{epsfig}
\usepackage{baltlat6}
\usepackage{array}
\usepackage{here}
\begin{document}
\ \
\vspace{0.5mm}
\setcounter{page}{000}

\titlehead{Baltic Astronomy, vol.\,N, NNN--NNN, 2015}

\titleb{CURRENT STAR FORMATION IN THE OUTER RINGS\\
AMONG EARLY-TYPE DISK GALAXIES}

\begin{authorl}
\authorb{I. P. Kostiuk}{1},
\authorb{O. K. Sil'chenko}{2}
\end{authorl}

\begin{addressl}
\addressb{1}{Special Astrophysical Observatory of the Russian Academy of Sciences,\\
Nizhnij Arkhyz 369167, Russia;\\
kostiuk@sao.ru}
\addressb{2}{Sternberg Astronomical Institute, Lomonosov Moscow State
University,\\  University av. 13, Moscow 119991, Russia;\\
olga@sai.msu.su}
\end{addressl}

\submitb{Received: ; accepted: }

\begin{summary} By using the ARRAKIS, the atlas of stellar rings in galaxies  (Comer\'on et al. 2014) 
based on data of the S4G survey, we have compiled a list of early-type, S0-Sb, disk galaxies
with outer stellar ring-like features (`pure' rings, R, or pseudorings, R$^{\prime}$). Current
star formation signatures within these features were searched for through the NUV-maps of
the galaxies provided by the ultraviolet space telescope GALEX. We have found that current
star formation, with the mean age of the young stellar population less than 200 Myr, is present 
in about a half of all `pure' rings; and within the pseudorings it is observed almost 
always. 
\end{summary}


%% \resthead is the RUNNING TITLE at top of the pages
\resthead{Star formation in outer rings of early-type galaxies}
{I. Kostiuk, O. Sil'chenko}

\sectionb{1}{INTRODUCTION}

Large-scale ring structures, both outer and inner ones, can be met in more than
half of all galactic disks (for the statistics -- see ARRAKIS, `Atlas of Resonance Rings
As Known In the S4G', Comer\'on et al. 2014). The rings described in the
ARRAKIS are purely stellar ones because at the wavelength of 4 mkm we see mostly
old stellar populations. On the contrary, in the near ultraviolet (NUV) spectral range
we can see mostly stars younger than a few hundred million years. And just it is in this
range that the public data of the space telescope GALEX (Gil de Paz et al. 2007) allow
to study morphology of nearby galaxies. Comer\'on (2014) has inspected the sample
of inner rings from the catalogue ARRAKIS by using the FUV maps from GALEX and 
narrow-band photometry data in the emission line H$\alpha$. His analysis reveals
that among early-type disk galaxies, S0-Sab, only 21\% ($\pm 3$\%) of the inner rings
are not seen in the FUV and so do not harbor current star formation. Consequently,
the dissipation time of the inner rings, in the frame of the resonance hypothesis on their
origin, exceeds 200 Myr that corresponds to about one orbital period in the central part
of the galaxies. We have undertaken similar analysis but we have applied it to the
outer rings in the early-type disks galaxies, to search for recent star formation there.

\sectionb{2}{THE SAMPLE}

We (Kostiuk \& Sil'chenko 2015) have compiled a list of 118 galaxies with outer ring-like
structures (`pure' rings R and pseudorings R$^{\prime}$) to study UV morphologies of
the outer rings in early-type disk galaxies; the catalogue of stellar ring-like structures 
ARRAKIS (Comer\'on et al. 2014) has been used as a source. The catalogue ARRAKIS is based on the 
3.6$\mu$m and 4.5$\mu$m data of the S4G survey (S4G$\equiv$ The Spitzer Survey of Stellar Structure 
in Galaxies, Sheth et al. 2010) which has covered a nearby galaxy sample limited by the following 
restrictions: the distance $D<40$~Mpc, the galactic latitude $\mid b \mid >30^{\circ}$, the integrated
magnitude $m_{\mbox{B,corr}} < 15.5$, the angular diameter $D_{25} > 1^{\prime}$. 
To determine surely parameters of ring-like structures, an additional restriction onto outer isophote 
ellipticity (and hence onto disk inclination to avoid strictly edge-on orientations) is needed; 
we apply the condition of $1-b/a < 0.5$ just as it was made in the ARRAKIS,
to derive statistical characteristics of the rings. 
For our study we have taken only outer rings, close to the optical
borders of the galactic disks, and only early morphological types of the disk galaxies, S0--Sb, 
where the ring structures are indeed frequent.

\sectionb{3}{STARFORMING RINGS}

We inspected our sample by using the public archive of
images (intensity maps from http://galex.stsci.edu/GR6/) of the space mission GALEX ; we explored
the near-ultraviolet band, NUV, 1770-2730\AA, where the light of B-stars
dominated. Because of their mass, the lifetime of these stars does not exceed 200~Myr, and so it is
the dating of recent star formation in the stellar rings. We estimated the mean NUV signal
in the outer rings and over background surrounding the galaxies. If the mean NUV signal in a ring
exceeds twice the background, we marked this galaxy as one having UV-radiation in its outer
ring-like structure (the notations are R$+$ and R$^{\prime} +$). 94 galaxies in our list are 
of SB or SAB type, so possessing bars; however the presence of bar has no influence 
onto the presence or absence of star formation in the outer ring. 84 galaxies of our list demonstrate
current star formation in their outer rings. The fraction of rings and pseudorings varies in galaxies 
according to their morphological types:
pseudorings are not found in S0 galaxies, instead `pure' outer rings are a typical feature of
S0s (60\%, according to Comer\'on et al. 2014); the early spirals, Sab--Sb, possess pseudorings
much more often than `pure' rings. Our results -- the frequency of current star formation
in the outer rings and pseudorings among the galaxies of different morphological types -- are shown
in Fig.~1. The fraction of both the `pure' rings and the pseudorings with current 
star formation rises when we pass from S0 to Sb through S0/a--Sa; but even in S0s where the fraction
of starforming outer rings is minimal, it reaches 56\%. Practically all spirals, Sab--Sb, have UV-signal
in the outer ring-like features. We specify three subclasses of UV morphology of the outer rings,
unclosed, clumpy, and in a filled disk, and among those the clumpy rings are the most frequent; moreover,
they are two times brighter in UV than the other subclasses.

\begin{figure}[!tH]
\vbox{
\centerline{\psfig{figure=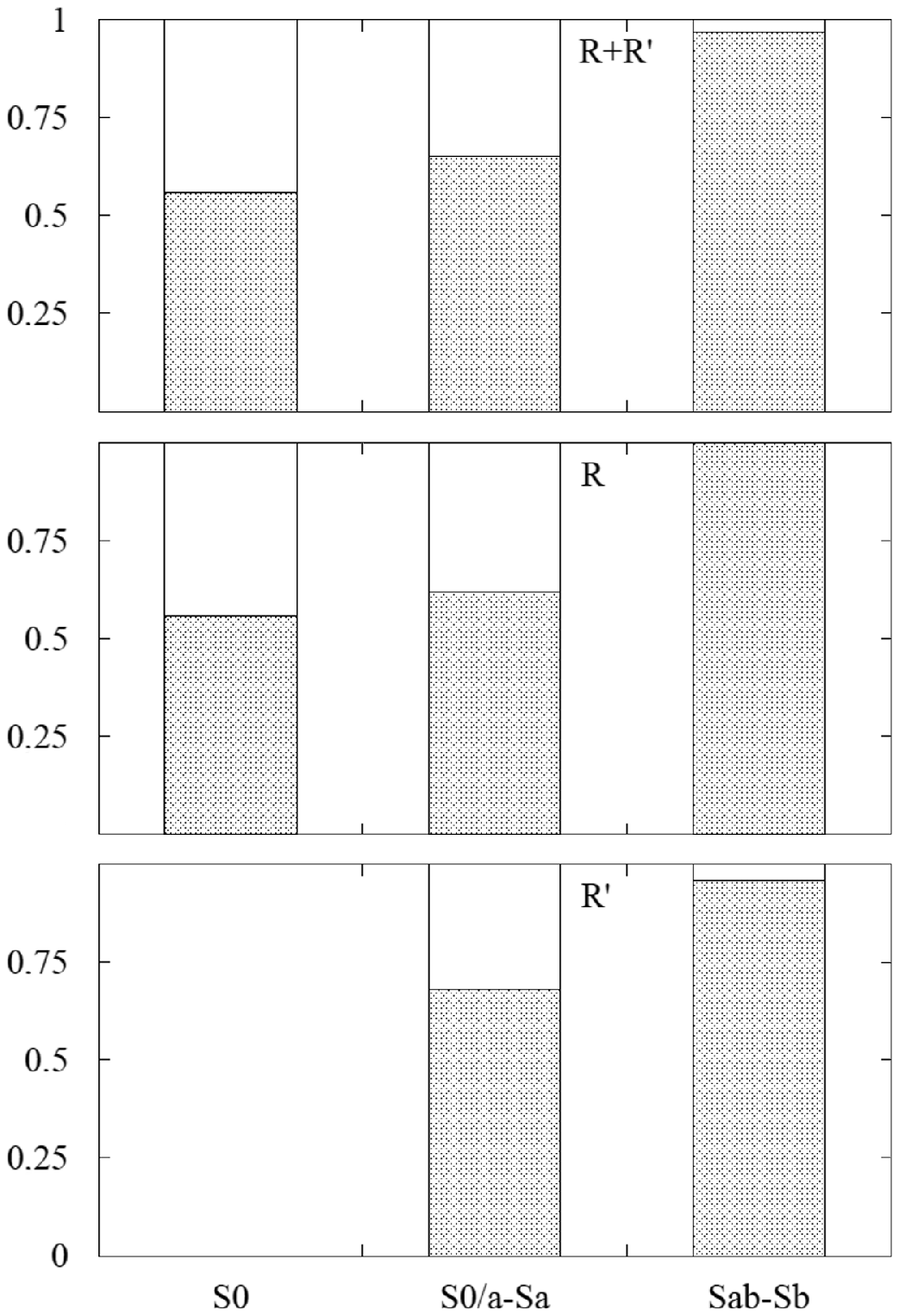,width=80mm}}
\vspace{1mm}
\captionb{1}
{The fraction of galaxies with UV-radiation in an outer ring ({\it shaded}) among
the outer stellar (4$\mu$m) rings in galaxies of different morphological types:
the upper row -- all the ring-like features, R$+$R$^{\prime}$, the middle row --
only `pure' rings, R, the bottom row -- pseudorings, R$^{\prime}$. Within every
type, a total galaxy number is normalized to unity.}
}
\end{figure}

%%%%%%%%%%%%%%%%%%%%%%%%%%%%%%%%%%%%%%%%

\sectionb{4}{DISCUSSION}

As noted by Sil'chenko et al. (2014), the nature of outer rings in disk galaxies
is controversial and is not understood yet. There are two popular scenaria 
of the outer ring origin -- 
resonance one and impact one. In the former scenario (Schommer \& Sullivan 1976,
Athanassoula et al. 1982, Buta \& Combes 1996), ring formation is
related to presence of a bar -- non-axisymmetric density perturbation 
in the center of a galaxy. The bar rotates as a rigid body, with the
angular velocity constant along the radius, and so it marks dynamically
certain radial zones where the differential rotation of the gas proceeds
in resonance with the bar. Near these resonances the orbits of gas clouds
crowd, the gas condenses, and necessary conditions for star formation ignition occur.
After the star formation proceeds, stellar rings emerge. Interestingly,
in the frame of this mechanism the first structures that form by gas cloud crowding
are pseudorings, with `pure' rings emerging later (Schwarz 1981, Byrd et al. 1994)); 
and according to our observational statistics the current star formation can be met 
more often in pseudorings than in `pure' rings that is in line with the theoretical results above. 
Also a theory of `manifolds' was proposed (Romero-G\'omez et al. 2007) which
stated the existence of persistent gas cloud orbits around stability points within a triaxial
(bar) potential. The impact mechanism -- the second most popular mechanism to form
rings in galactic disks -- developped by Freeman \& de Vaucouleurs (1974),
Theys \& Spiegel (1976), Few \& Madore (1986), Athanassoula et al. (1997), implies 
drop of a satellite from highly inclined orbit onto galactic disk
near the center. Such impact provokes a circular gas density wave running
outward through the disk. As a result of gas compression, at some radius, 
again, star formation may ignite, and a stellar ring may form. 
In the case of a pure stellar disk, the impact can generate transient
stellar density enhancement looking as a ring moving outward (Wu \& Jiang 2012).
A third possibility to form an outer in-plane ring -- to accrete cold gas from 
outside, as a result of tidal interaction or from a cosmological filament -- 
was mentioned by Buta \& Combes (1996) and by Byrd et al. (1994) but it was not 
discussed exhaustively yet.

Meantime Pogge \& Eskridge (1993) who searched for star formation in HI-rich S0 galaxies,
noted that, firstly, star formation in S0s was always organized in ring-like
structures, and secondly, the occurence of star formation (it was found
in a half of gas-rich S0s) did not depend on the amount of gas. A hypothesis was then
formulated that star formation in gas-rich S0s had to be drived not by intrinsic
gravitational instability, but by some kinematical effect.
Indeed, surface density of HI in outer disks of early-type galaxies was often
below the Kennicutt's (1989) threshold, and e. g. Noordermeer et al. (2005) noted that
star formation in the outer ring of NGC~7217 could not proceed because the gas had to be
stable. However, observations reveal star formation sites in this low-density gas.
Some external triggering, like outer cold gas supply and shock
compression due to accretion, seems to be rather necessary.

It is known that cold gas distribution in early-type galaxies where the gas is found
can be even more extended than in spirals -- regular HI disks in S0s may reach up to 200 kpc
in diameter (Oosterloo et al. 2007). The statistics by Afanas'ev \& Kostyuk (1988)
based on surface photometry showed that galaxies with outer rings
possess more extended stellar disks than galaxies without rings. We think that it is due
to disk growth inside-out through the stimulated star formation in the outer cold-gas
rings accreted from outside.

\thanks{The work is based on data of the space mission GALEX
via The MultiMission Mikulski Archive for Space Telescopes (MAST)
at Space Telescope Science Institute.
During the data analysis we use the NASA/IPAC
Extragalactic Database (NED) which is operated by the Jet Propulsion
Laboratory, California Institute of Technology, under contract with
the National Aeronautics and Space Administration. 
The study of disk galaxy
structureû and evolution is financed by the grant of the Russian Science Foundation
14-22-00041.}

\References

\refb Afanas'ev V. L. \& Kostyuk I. P. 1988, Astrofizika, 29, 531

\refb Athanassoula E., Bosma A., Creze M., Schwarz M. P. 1982, A\& A, 107, 101

\refb Athanassoula E., Puerari I., Bosma A. 1997, MNRAS, 286, 284

\refb Byrd G., Rautiainen P., Salo H., Buta R., Crocker D. A. 1994, AJ, 108, 476

\refb Buta R. \& Combes F. 1996, Fundamentals of Cosmic Physics, 17, 95

\refb Comer\'on S., Salo H., Laurikainen E. et al. 2014, A\&A, 562, 121

\refb Comer\'on S. 2013, A\&A, 555, L4

\refb Few J. M. A. \& Madore B. F. 1986, MNRAS, 222, 673

\refb Freeman K. C. \& de Vaucouleurs G. 1974, ApJ, 194, 569

\refb Gil de Paz A., Boissier S., Madore B. F. et al. 2007, ApJS, 173, 185

\refb Kennicutt Jr. R. C. 1989, ApJ, 344, 685

\refb Kostiuk I. P. \& Sil'chenko O. K. 2015, Astrophys. Bull., 70, 280

\refb Noordermeer E., van der Hulst J. M., Sancisi R., et al. 2005, A\&A, 442, 137

\refb Oosterloo T., Morganti R., Sadler E. et al. 2007, A\&A, 465, 787

\refb Pogge R. W. \& Eskridge P. B. 1993, AJ, 106, 1405

\refb Romero-G\'omez M., Athanassoula E., Masdemont J.J., et al. 2007, A\&A, 472, 63

\refb Schommer R. A., Sullivan W. T. III 1976, Aph. Lett., 17, 191

\refb Sheth K., Regan M., Hinz J. L. et al. 2010, PASP, 122, 1397

\refb Silchenko O., Ilyina M., Katkov I. 2014, Baltic Astronomy, 23, 279

\refb Schwarz M. P. 1981, ApJ, 247, 77

\refb Theys J. C. \& Spiegel E. A. 1976, ApJ, 208, 650

\refb Wu Y. T. \& Jiang I. G. 2012, ApJ, 745, 105

\end{document}